\documentstyle[prc,aps,graphicx,dcolumn,bm]{revtex}

\begin{document}

\twocolumn[\hsize\textwidth\columnwidth\hsize  
\csname @twocolumnfalse\endcsname              

\draft

\author{P. Boutachkov$^{a}$, A. Aprahamian$^{a}$, Y. Sun$^{b,c}$,
J.A. Sheikh$^{d}$, S. Frauendorf$^{a}$}
\address{$^{a}$ Department of Physics, University of Notre Dame, Notre Dame, IN
46556, U.S.A.}
\address{$^{b}$ Department of Physics and Astronomy, University of Tennessee,
 Knoxville, TN 37996, U.S.A.}
\address{$^{c}$ Department of Physics, Xuzhou Normal University, Xuzhou, Jiangsu 221009,
 P.R. China}
\address{$^{d}$ Physik-Department, Technische Universit\"at
 M\"unchen, D-85747 Garching, Germany}

\title{In-Band and Inter-Band $B(E2)$ Values within the Triaxial Projected Shell Model}

\maketitle

\begin{abstract}
The Triaxial Projected Shell Model (TPSM) has been successful in
providing a microscopic description of the energies of
multi-phonon vibrational bands in deformed nuclei. We report here
on an extension of the TPSM to allow, for the first time,
calculations of $B(E2)$ values connecting $\gamma$- and
$\gamma\gamma$-vibrational bands and the ground state band. The
method is applied to $^{166,168}$Er. It is shown that most of the
existing $B(E2)$ data can be reproduced rather well, thus
strongly supporting the classification of these states as
$\gamma$-vibrational states. However, significant differences
between the data and the calculation are seen in those $B(E2)$
values which involve odd-spin states of the $\gamma$-band.
Understanding these discrepancies requires accurate experimental
measurements and perhaps further improvements of the TPSM.
\end{abstract}

\pacs{PACS: 21.60.Cs, 
21.10.Re,
21.10.Ky,
27.70.+q}
]

\bigskip
Recently two-phonon
$\gamma\gamma$-vibrational bands have been identified 
in a number of nuclei \cite{Wu94,DP168Er,GAMMA166Er}, where 
pronounced anharmonicities have been observed in the  vibrational spectrum. A
microscopic description of the energies and transition
probabilities of two-phonon vibrational excitations remains
a challenge to nuclear models.
The Triaxial Projected Shell Model \cite{TPSM1,TPSM2} 
is a new microscopic, fully quantum-mechanical
model with a unified treatment of the vibrational and rotational
states.
In the TPSM approach, one introduces triaxiality in the deformed
basis and performs exactly three-dimensional angular momentum
projection \cite{TPSM1}. In this way, the deformed vacuum state is
much enriched by allowing all possible $K$-components.
Diagonalization mixes these components, and various excited bands 
emerge \cite{TPSM2} besides the ground state (g.s.) band ($K=0$).
The excited $K=2$ band describes the one-phonon
$\gamma$-vibrational band; and the excited band with $K=4$
accounts for  the  two-phonon $\gamma\gamma$-band. The
observed anharmonicities in the energies of multi-phonon
vibrational bands occur quite naturally from the TPSM without
including additional ingredients in the model \cite{TPSM2}. 
The
TPSM has recently been applied also to the study of transition
quadrupole moments in the g.s. bands of $\gamma$-soft nuclei and
the magnetic dipole properties of the $\gamma$-vibrational states
\cite{Sheikh2001,Sun2002}.
However, in-band and inter-band
E2-transitions, which are important quantities in supporting
classification of states, have not been studied yet.

The purpose of this paper is to report a new
extension of the TPSM which allows the
calculation of $B(E2)$ values for transitions connecting the
g.s. band, the single $\gamma$, and double $\gamma\gamma$
vibrational bands. We have searched the entire rare-earth region 
for experimental inter-band $B(E2)$ values that allow 
a comparison with our calculations.  Data on absolute
$B(E2)$ values are sparse and  only in few cases they are known
for more than one or two members of the  $\gamma$-band. 
In this paper we study $B(E2)$ values for
$^{166,168}$Er, which are the best cases. 
These nuclei exhibit well established
double-phonon excitations \cite{DP168Er,GAMMA166Er}.
$^{168}$Er is one of the most extensively studied nuclei in this
mass region with several measured $B(E2)$ values for members of
the $\gamma$-band \cite{N168Er} and $\gamma\gamma$-band \cite{GAMMA166Er}.

In the TPSM, one calculates the $\gamma$-vibrational states by
building a shell model space truncated in a triaxially deformed
basis. This is done by an exact three-dimensional
angular-momentum projection of the $\gamma$-deformed Nilsson +
BCS basis $\left| \Phi \right\rangle $. The Nilsson Hamiltonian
is:
\begin{equation}
\hat{H}=\hat{H}_0-\frac 23\hbar \omega \left[ \epsilon
\hat{Q}_0+\epsilon
^{\prime}\frac{\hat{Q}_{+2}+\hat{Q}_{-2}}{\sqrt{2}}\right]
\label{Nils}
\end{equation}
where $\hat{H}_0$ is the spherical single-particle Hamiltonian
with inclusion of the appropriate spin-orbit forces parameterized
by Nilsson {\it et al.} \cite{NilLS}. The axial
and the triaxial parts of the Nilsson potential  in Eq.
(\ref{Nils}) contain the parameters $\epsilon$ and $\epsilon ^{\prime}$
respectively, which are related to 
the conventional triaxiality  parameter by $\gamma
= \tan(\frac{\epsilon ^{\prime}}\epsilon )$. The rotational invariant
two-body Hamiltonian
\begin{equation}
\hat{H}=\hat{H}_0-\frac \chi2 \sum_\mu \hat{Q}_\mu ^{+}\hat{Q}_\mu -G_M\hat{%
P}^{+}\hat{P}-G_Q\sum_\mu \hat{P}_\mu ^{+}\hat{P}_\mu
\label{Hamilt}
\end{equation}
is diagonalized in the TPSM basis:

$\left\{ \hat{P}_{MK}^I\left| \Phi \right\rangle ,0\leq K\leq
I\right\}$. 
The  solutions take the form
\begin{equation}
\left| \Psi^\sigma_{IM}\right\rangle =\sum_{0\leq K\leq I}f^\sigma_{IK} \hat{P}%
_{MK}^I\left| \Phi \right\rangle ,
\label{Equat}
\end{equation}
where $\sigma$ specifies the states with the same angular
momentum $I$.  The strength
of the monopole and quadrupole pairing forces is set by $G_M$ and $G_Q$ in 
Eq. (\ref{Hamilt}), where $G_M=\left[ G_1\pm
G_2\frac{N-Z}A\right] /A$ with $"+"$ for protons and $"-"$ for
neutrons. We use $G_1=20.12$, $G_2=13.13$ and $G_Q=0.16G_M$,
which are the same values used in previous calculations
\cite{TPSM1,TPSM2,PSM}. The $QQ$-force strength $\chi $ is
determined such that it holds a self-consistent relation
with the quadrupole deformation $\epsilon $ \cite{PSM}.

Once the Hamiltonian is diagonalized in the TPSM basis, the
eigenfunctions are used to calculate the electric quadrupole
transition probabilities
\[
B(E2: (I_i, K_i)\rightarrow (I_f, K_f))=\frac 1{2I_i+1}\left|
\left\langle \Psi^{K_f}_{I_f} || \hat Q_2 || \Psi^{K_i}_{I_i}
\right\rangle \right| ^2
\]
between an initial state $(I_i, K_i)$ and a final states $(I_f,
K_f)$. The explicit expression for the reduced matrix element in
the projected basis can be found in Ref. \cite{Sheikh2001}. Note
that we now use $K$ instead of $\sigma$ to specify states with the
same angular momentum $I$. According to Eq. (\ref{Equat}),
 $K$ is not a good quantum number. However, it has been  shown
\cite{TPSM2} that in these well-deformed nuclei, $K$-mixing is
rather weak. Thus, we use $K$  to denote bands keeping  the 
familiar convention. In
the calculation, we use the standard effective charges of
1.5$e$ for protons and 0.5$e$ for neutrons.

In the present calculation
, the parameters $\epsilon $ and $\epsilon ^{\prime}$ are
 considered as adjustable. 
 For the deformation parameter $\epsilon$
the experimental value 0.320 \cite{RaEPS} is used, which means 
it is chosen such that the experimental value of 
$B(E2:2^{+}_{K=0}\rightarrow 0^{+}_{K=0})$ is approximately reproduced. 
In previous applications, the calculated deformation parameters \cite{NilEPS}
 were used, which means in the present case $\epsilon=0.273$. Besides giving
a better scale for the $B(E2)$ values,  
the use of the experimental $\epsilon$  
slightly better reproduces the energy levels
\cite{PB}. Except for the overall scale,  
$B(E2)$ values are not very sensitive to the moderate changes
of $\epsilon$, in
particular the $B(E2)$ values of the $\gamma$-vibrational states.
The triaxiality parameter $\epsilon ^{\prime}$ is  chosen so that 
the calculated energy of the $K=2$ band-head  reproduces the measured value. 
For $^{168}$Er, we find  $\epsilon ^{\prime}=0.125$. 

\begin{figure}[htb]
\vspace{0.2cm}
{\rotatebox{0}{\includegraphics[scale=0.35]{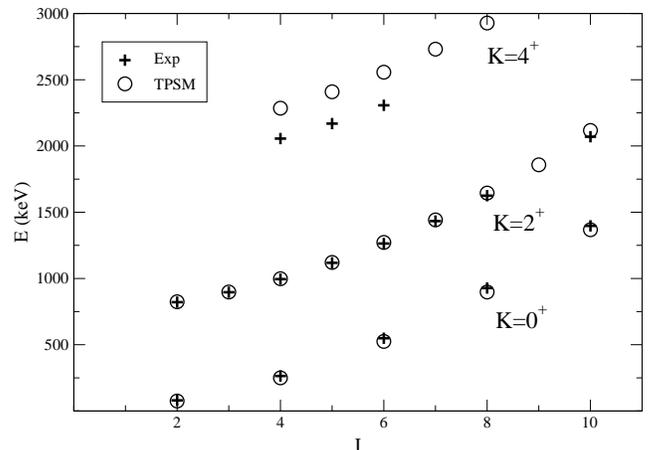}}}
\caption{The energies \protect\cite{DP168Er,N168Er,DAV} of levels within the g.s., $\gamma$-, and
$\gamma\gamma$-bands in $^{168}$Er compared with those
calculated within the TPSM as a function of spin I. }
\label{figure.1}
\end{figure}

The experimental and calculated energies for $^{168}$Er 
are compared in  Fig. \ref{figure.1}. 
It can be seen that all the energy levels in
the g.s. band, the $\gamma$-band, and the $K=4$
$\gamma\gamma$-band are well described within the TPSM. Without
introducing additional ingredients, the model gives 
anharmonicities in the energies of the $\gamma\gamma$ vibrational
bands, which are bigger in comparison with experiment.

\begin{figure}[htb]
{\rotatebox{-90}{\includegraphics[scale=0.5]{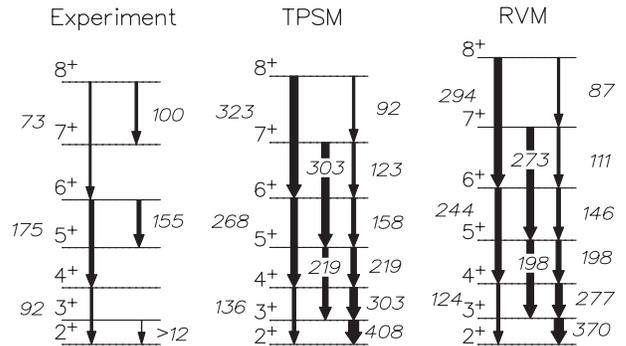}}}
\vspace{0.2cm}
\caption{Comparison of experimental in-band B(E2) values for the
K$^{\pi}=2^+$ $\gamma$-band with calculations of the TPSM and the
RVM in $^{168}$Er. The thickness of the arrows is proportional to
the magnitude of the transition probabilities in W.u.}
\label{figure.2}
\end{figure}

The calculated $B(E2)$ values from the TPSM are not far from the ones for a
 rotor coupled to a harmonic vibrator (see e.g. \cite{RVM}).
This familiar limit  is included in Fig. \ref{figure.2}, where the  $B(E2)$ values for transitions within
the $\gamma$-band of   $^{168}$Er are compared with
experimental data from Ref.\cite{DP168Er,N168Er,DAV,g-Er168}. Table \ref{table.1} and Fig. \ref{figure.2}
compare the complete set of $B(E2)$ values. 
Except for some particular
transitions that we shall discuss below, most of the theoretical
$B(E2)$ values appear to agree with the measured values. In general,
the in-band transition probabilities are two order of magnitude
stronger than the inter-band transitions. 
The $B(E2)$ transitions from the $\gamma\gamma$-band are
quite well reproduced in TPSM. 

There may be a discrepancy  between the calculated and
measured $B(E2)$ values involving the odd-spin $3^+_{K=2}$ state of the
$\gamma$-band. For example, the experimental
$B(E2:3^+_{K=2}\rightarrow2^+_{K=2})$ value is $>12$ $W.u.$ while
the calculated value from the TPSM gives 408 $W.u.$ Similarly,
the calculated $B(E2:3^+_{K=2}\rightarrow2^+_{K=0})$ is $5$
$W.u.$, which is by an order of magnitude larger than the
experimental lower limit of $>0.2$ $W.u$. The
$B(E2:3^+_{K=2}\rightarrow4^+_{K=0})$  is also an order
of magnitude off the limit. 
The experimental
values are lower limits because  there is only an upper
limit known for the lifetime \cite{odd-spin} of the $3^+_{K=2}$ state.
The $3^{+}_{K=2}$ state is described as a
rotational state built on a $\gamma$-vibration. Therefore all
models that use this picture will give a large probability for the
$3^{+}_{K=2}\rightarrow 2^{+}_{K=2}$ transition. Accordingly, 
the large experimental $B(E2)$ value for the
$4^{+}_{K=2}\rightarrow 2^{+}_{K=2}$ transition in the $\gamma$-band is very
well described by the theory.  
Clearly a more accurate lifetime
measurement of the $3^+_{K=2}$ state  
$^{168}$Er is desirable in order to settle the question, whether there is a
discrepancy between theory and experiment for transitions involving the
odd-spin states or not.

One expects that collective levels with energies
larger than the pairing gap $2\Delta$  
 are to some extent mixed with the
two-quasiparticle excitations. The present
calculations do not explicitly include excited quasiparticle configurations
based on the $\gamma$-deformed basis.
The experimental $B(E2)$ values are well reproduced 
for states that are expected to weakly mix with two-quasiparticle excitations. The only exception is the 
 $3^+_{K=2}$ state, which is inconclusive.
The states of the $\gamma\gamma$-band lie in the energy region where
mixing with the two-quasiparticle states should become important.
The calculated $B(E2)$ value for the
$4^{+}_{K=4}\rightarrow 2^{+}_{K=2}$ transition is about four times 
and the one for the $4^{+}_{K=4}\rightarrow 3^{+}_{K=2}$ is about
three times larger than in experiment. A substantial admixture of a
two-quasiparticle states into the collective $\gamma\gamma$-vibration would
reduce this $B(E2)$ value. Such a mixing may also account for the 
deviation between the experimental and calculated energies of the 
$\gamma\gamma$-vibration.
The other  significant discrepancy is associated with the
$8^{+}_{K=2}$ level at $1625$ $keV$.  The calculated 
$B(E2:8^{+}_{K=2} \rightarrow 6^{+}_{K=2})$ is 323 $W.u.$, which is
about 4.5 times larger than the measured value. At this excitation
energy and angular momentum,
single particle excitations are expected mix with the collective states,
leading to crossing between collective and two-quasiparticle bands.
The $B(E2:8^{+}_{K=2} \rightarrow 10^{+}_{K=0})$ shown
in Table \ref{table.1} has a
calculated value of 1.4 $W.u$. One would expect that the 
single particle effects will tend to further reduce this number, but the
experimental value is $120\pm 50$ $W.u.$ The reason for the increased collectivity
is not clear. This is an indication
that at high spins the $\gamma$-band behaves differently from what
is expected for  a collective excitation.

\begin{figure}[htb]
\vspace{0.2cm}
{\rotatebox{0}{\includegraphics[scale=0.35]{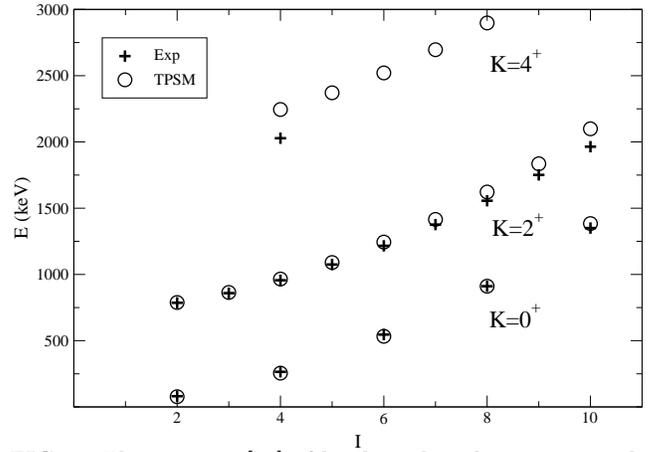}}}
\caption{The energies \protect\cite{NDS166Er} of levels within the g.s., $\gamma$-, and
$\gamma\gamma$-bands in $^{166}Er$ compared with calculated values from the TPSM as a function of spin I.
.}
\label{figure.3}
\end{figure}

We also calculated energies and B(E2)
values for $^{166}$Er. Here, we use $\epsilon =0.324$
\cite{RaEPS} and $\epsilon^{\prime}=0.126$; all other
parameters are the same as for $^{168}$Er. The experimental data
for the g.s. band, $\gamma$-band energy, life times and
intensities of the $\gamma$ transitions of interest are taken
from Ref. \cite{GAMMA166Er,NDS166Er}. The data for the $\gamma\gamma$-band
is taken from Ref. \cite{GAMMA166Er}. Fig. \ref{figure.3} and Table \ref{table.2} show the
level energies for the g.s., $\gamma$-, $\gamma \gamma$-bands,
and the B(E2) values for $^{166}$Er, respectively.

This calculation leads to the same conclusion as in the
$^{168}$Er case: The TPSM describes the energies of the g.s. band
and the $\gamma$-band well. The energies of the $\gamma
\gamma$-band are reproduced quite well for $^{166}$Er (see Fig.
\ref{figure.3}). In fact, one can argue that the
$4^{+}_{K=4}$ state wave function in $^{166}$Er is more collective
than in $^{168}$Er because the $B(E2)_{TPSM}$ value of $12$
$W.u.$ is closer to the experimental one of $7.4$ $W.u$. For the 
transitions from the odd-spin states of the $\gamma$-band there is a similar 
inconclusive situation as in $^{168}$Er. In $^{166}$Er,
only the upper limit on the lifetime of the $5^+_{K=2}$ state is known
experimentally. The calculated $B(E2:5^{+}_{K=2} \rightarrow
3^{+}_{K=2})$ value is $223$ $W.u.$ while the experimental limit
is $>14$ $W.u$. There is an order of magnitude difference between the 
theoretical 
$B(E2:5^{+}_{K=2} \rightarrow 4^{+}_{K=0})$ and $B(E2:5^{+}_{K=2}
\rightarrow 6^{+}_{K=0})$ values and the experimental limits.

In conclusion, 
the Triaxial Projected Shell Model
has been successful in describing the 
experimental level energies for the g.s., the $\gamma$, and the
$\gamma\gamma$-bands with their inherent anharmonicities. We have 
calculated for the first time, $B(E2)$ values for inter-band transitions 
between the g.s., $\gamma$-, and
$\gamma\gamma$-bands in $^{166,168}$Er.  
Most of the calculated $B(E2)$ values well agree 
with the available experimental data.
Only lower limits for  $B(E2)$ values associated with the odd-spin 
members of the $\gamma$-band can be derived from the available data. 
More accurate lifetime measurements are necessary for a stringent test of the
theory. The deviations between calculated and experimental 
$B(E2)$ values seem to point to the 
inclusion of two-quasiparticle admixtures in the
collective excitations. Hence, it appears 
necessary to explicitly include excited quasi-particle configurations
into the Triaxial Projected Shell Model in order to achieve an understanding of
the nature of vibrational states in deformed nuclei where fragmentation of collectivity 
among quasiparticle excitations is expected to play an important role.

Research on this topic was supported by the National Science
Foundation under the  contract 99-01133 and the Department of Energy under the
grant DE-FG02-95ER40934.

\begin{table}[h]
\caption{Comparison of all known experimental in-band and
inter-band $B(E2)$ values (associated errors in parenthesis) and calculated ones in $W.u.$ for
$^{168}$Er.
K=4$^{+}$ lifetimes from Ref.\protect\cite{DP168Er}, K=0$^{+}$, and K=2$^{+}$ lifetimes and B(E2) values from
Ref.\protect\cite{N168Er} and all the references therein.\\
$^{*}$ B(E2) value from Ref.\protect\cite{g-Er168}; the calculated axial rotor value is $336$ $W.u$.\\
$^{**}$ B(E2) values calculated from lifetimes in Ref.\protect\cite{DP168Er}.
}

\begin{tabular}{|c|c|c|}
\hline
$(I,K)_i\rightarrow (I,K)_f$ & $B(E2)_{\exp} (W.u.)$ & 
$B(E2)_{TPSM}(W.u.)$ \\ \hline\hline
$(2,0)_i\rightarrow (0,0)_f$ & 207 (\textit{6)} & 228.6 \\ \hline
$(4,0)_i\rightarrow (2,0)_f$ & 318 (\textit{12)} & 326.9 \\ \hline
$(6,0)_i\rightarrow (4,0)_f$ & 440* (30) & 361.2 \\ \hline
$(8,0)_i\rightarrow (6,0)_f$ & 350 (\textit{20)} & 380.0 \\ \hline
$(10,0)_i\rightarrow (8,0)_f$ & 302 (\textit{21)} & 393.0 \\ \hline\hline
$(2,2)_i\rightarrow (0,0)_f$ & 4.80 (\textit{17) } & 2.7 \\ \hline
$(2,2)_i\rightarrow (2,0)_f$ & 8.5 (\textit{4)} & 4.5 \\ \hline
$(2,2)_i\rightarrow (4,0)_f$ & 0.62 (\textit{4)} & 0.3 \\ \hline
$(3,2)_i\rightarrow (2,0)_f$ & $>0.19$ & 4.9 \\ \hline
$(3,2)_i\rightarrow (4,0)_f$ & $>0.13$ & 2.7 \\ \hline
$(4,2)_i\rightarrow (2,0)_f$ & 1.7 (\textit{4)} & 1.3 \\ \hline
$(4,2)_i\rightarrow (4,0)_f$ & 8.7 (\textit{18)} & 5.5 \\ \hline
$(4,2)_i\rightarrow (6,0)_f$ & 1.13 (\textit{25)} & 0.7 \\ \hline
$(5,2)_i\rightarrow (4,0)_f$ &  & 3.9 \\ \hline
$(5,2)_i\rightarrow (6,0)_f$ &  & 3.7 \\ \hline
$(6,2)_i\rightarrow (4,0)_f$ & 0.78 (\textit{19)} & 0.8 \\ \hline
$(6,2)_i\rightarrow (6,0)_f$ & 6.4 (\textit{16)} & 5.7 \\ \hline
$(6,2)_i\rightarrow (8,0)_f$ & 2.4 (\textit{7)} & 1.1 \\ \hline
$(7,2)_i\rightarrow (6,0)_f$ &  & 3.3 \\ \hline
$(7,2)_i\rightarrow (8,0)_f$ &  & 4.4 \\ \hline
$(8,2)_i\rightarrow (6,0)_f$ & 1.3 (\textit{6)} & 0.5 \\ \hline
$(8,2)_i\rightarrow (8,0)_f$ & 1.8 (\textit{8)} & 5.7 \\ \hline
$(8,2)_i\rightarrow (10,0)_f$ & 120 (\textit{50)} & 1.4 \\ \hline\hline
$(4,4)_i\rightarrow (2,2)_f$ & 3.4 (\textit{19)} & 11.9 \\ \hline
$(4,4)_i\rightarrow (3,2)_f$ & 2.2 (\textit{13)} & 7.1 \\ \hline
$(4,4)_i\rightarrow (4,2)_f$ & 1.7** (\textit{9)} & 2.7 \\ \hline
$(4,4)_i\rightarrow (5,2)_f$ & 0.7** (\textit{3)} & 0.6 \\ \hline
$(4,4)_i\rightarrow (6,2)_f$ & 2.0 (\textit{13)} & 0.1 \\ \hline
$(5,4)_i\rightarrow (3,2)_f$ & 5 (\textit{5)} & 7.7 \\ \hline
$(5,4)_i\rightarrow (4,2)_f$ & 4\textit{\ (3)} & 8.6 \\ \hline
$(5,4)_i\rightarrow (5,2)_f$ & 1.8 (\textit{15)} & 4.6 \\ \hline
$(5,4)_i\rightarrow (6,2)_f$ & 0.8 (\textit{7)} & 1.3 \\ \hline
$(5,4)_i\rightarrow (7,2)_f$ & 7 (\textit{6)} & 0.2 \\ \hline
\end{tabular}

\label{table.1}
\end{table}

\begin{table}[h]
\caption{Comparison of all known experimental in-band and
inter-band $B(E2)$ values (associated errors in parenthesis) and calculated ones in $W.u.$ for
$^{166}$Er.\\
$^{*}$ $B_{exp}(E2)$ is calculated as an upper limit assuming 100\% E2.\\
$^{**}$ Data from Ref.\protect\cite{GAMMA166Er}.}

\begin{tabular}{|c|c|c|}
\hline
$(I,K)_i\rightarrow (I,K)_f$ & $B(E2)_{exp}(W.u.)$ & 
$B(E2)_{TPSM}(W.u.)$ \\ \hline\hline
$(2,0)_i\rightarrow (0,0)_f$ & 214 (\textit{10)} & 231.6 \\ \hline
$(4,0)_i\rightarrow (2,0)_f$ & 311 (\textit{20)} & 331.3 \\ \hline
$(6,0)_i\rightarrow (4,0)_f$ & 347 (\textit{45)} & 366.2 \\ \hline
$(8,0)_i\rightarrow (6,0)_f$ & 365 (\textit{50)} & 385.5 \\ \hline
$(10,0)_i\rightarrow (8,0)_f$ & 371 (\textit{46)} & 399.1 \\ \hline\hline
$(3,2)_i\rightarrow (2,2)_f$ &  & 414.0 \\ \hline
$(4,2)_i\rightarrow (2,2)_f$ &  & 137.8 \\ \hline
$(4,2)_i\rightarrow (3,2)_f$ &  & 306.8 \\ \hline
$(5,2)_i\rightarrow (3,2)_f$ & $>14$ & 222.0 \\ \hline
$(5,2)_i\rightarrow (4,2)_f$ & $>18$* & 221.6 \\ \hline\hline
$(2,2)_i\rightarrow (0,0)_f$ & 5.5 (\textit{4)} & 2.8 \\ \hline
$(2,2)_i\rightarrow (2,0)_f$ & 9.7 (\textit{7)} & 4.7 \\ \hline
$(2,2)_i\rightarrow (4,0)_f$ & 0.67 (\textit{5)} & 0.3 \\ \hline
$(5,2)_i\rightarrow (4,0)_f$ & $>0.4$ & 3.8 \\ \hline
$(5,2)_i\rightarrow (6,0)_f$ & $>0.6$ & 4.1 \\ \hline\hline
$(4,4)_i\rightarrow (2,2)_f$ & 7.4**  (\textit{2.5)} & 12.1 \\ \hline
$(4,4)_i\rightarrow (3,2)_f$ &  & 8.7 \\ \hline
$(4,4)_i\rightarrow (4,2)_f$ &  & 2.9 \\ \hline
$(4,4)_i\rightarrow (5,2)_f$ &  & 0.7 \\ \hline
\end{tabular}

\label{table.2}
\end{table}
\end{document}